\documentclass[a4paper,12pt,twoside]{article}
\usepackage{authblk}
\usepackage{a4wide}
\usepackage[utf8]{inputenc}
\usepackage{graphicx}
\usepackage{amssymb,amsmath,textcomp}
\usepackage{natbib}
\usepackage{geometry}
\usepackage{float}
\usepackage{color}
\usepackage{times}
\usepackage{hyperref}
\definecolor{Brown}{rgb}{0.647,0.165,0.165}
\definecolor{NavyBlue}{rgb}{0.0,0,0.5}
\definecolor{Burgundy}{rgb}{0.5,0.0,0.125}
\hypersetup{
    breaklinks=true,
    colorlinks=true,
    citecolor={Burgundy},
    linkcolor={NavyBlue},
    urlcolor={Brown}
}
\usepackage{parskip}
\graphicspath{{./}{figs/}}
\usepackage[explicit,compact]{titlesec}
\titlespacing{\section}{0pt}{1ex}{1ex}
\titleformat{\section}{\bfseries\large\color{NavyBlue}}{\thesection}{10pt}{\MakeUppercase{#1}}%

\usepackage{newtxtext,newtxmath}

\usepackage[T1]{fontenc}

\DeclareRobustCommand{\VAN}[3]{#2}
\let\VANthebibliography\thebibliography
\def\thebibliography{\DeclareRobustCommand{\VAN}[3]{##3}\VANthebibliography}
\DeclareRobustCommand{\DE}[3]{#2}
\let\DEthebibliography\thebibliography
\def\thebibliography{\DeclareRobustCommand{\DE}[3]{##3}\DEthebibliography}

\usepackage{graphicx}	
\usepackage{amsmath}	
\usepackage{pdflscape}
\usepackage{rotating}

\PassOptionsToPackage{hyphens}{url}
\usepackage[all]{hypcap}
\usepackage{color}
\definecolor{Brown}{rgb}{0.647,0.165,0.165}
\definecolor{NavyBlue}{rgb}{0.0,0,0.5}
\definecolor{Burgundy}{rgb}{0.5,0.0,0.125}
\hypersetup{
    breaklinks=true,
    colorlinks=true,
    citecolor={Burgundy},
    linkcolor={NavyBlue},
    urlcolor={Brown}
}
\usepackage{multirow}
\def\apj{Astrophys.\ J.}

\def\mnras{Mon.\ Not.\ R.\ Astron.\ Soc.}
\def\aap{Astron.\ Astrophys.}
\def\apjl{Astrophys.\ J.\ Lett.}
\def\aapr{Astron.\ Astrophys.\ Rev.}

\def\physrep{Phys.\ Rep.}
\def\pre{Phys.\ Rev.\ E}
\def\prl{Phys.\ Rev.\ Lett.}

\def\nat{Nature}

\def\araa{Ann.\ Rev.\ Astron.\ Astrophys.}

\newcommand{\Mach}{\mathcal{M}}      
 
\renewcommand{\vec}[1]{\boldsymbol{#1}}	

\newcommand{\cm}{{\rm cm}}    
\newcommand{\km}{{\rm km}}    
\newcommand{\pc}{{\rm pc}}     
\newcommand{\kpc}{{\rm kpc}}  

\newcommand{\g}{{\rm g}}      

\newcommand{\s}{{\rm s}}      
\newcommand{\Myr}{{\rm Myr}} 
\newcommand{\Gyr}{{\rm Gyr}}  

\newcommand{\muG}{\mu{\rm G}} 

\newcommand{\K}{{\rm K}}      

\newcommand{\Emag}{E_{\rm mag}}
\newcommand{\Ekin}{E_{\rm kin}}

\newcommand{\cold}{T < 10^3~{\rm K}}
\newcommand{\warm}{T \ge 10^3~{\rm K}}
\newcommand{\whole}{T \ge 0~{\rm K}}

\newcommand\Fig[1]{{\color{NavyBlue} Fig.~\ref{#1}}}
\newcommand\Sec[1]{{\color{NavyBlue} Sec.~\ref{#1}}}

\graphicspath{{./}{figs/}}

\usepackage{fancyhdr}

\fancypagestyle{firstpage}{
  \lhead{Conference Proceedings for Cosmic Magnetism in the Pre-SKA Era \\ May 27 -- 31, 2024, Inamori Kaikan, Kagoshima University \\ Kagoshima, Japan}
  \rhead{}
}
\title{\vspace{-0.25cm} \huge{Turbulent Dynamo and Magnetic Fields in the Multiphase Interstellar Medium}}
\author{\Large{Amit Seta}\thanks{E-mail:  \href{mailto:amit.seta@anu.edu.au}{amit.seta@anu.edu.au}}} \affil{\large{Research School of Astronomy and Astrophysics,  \\  Australian National University, \\ Canberra, ACT 2611, Australia}} 
\usepackage{datetime}
\newdate{date}{20}{09}{2024}
\date{\today}

\begin{document}
\maketitle
\thispagestyle{firstpage}

\begin{abstract}
\noindent Even though the interstellar medium (ISM) of star-forming galaxies has been known to have a multiphase structure (broadly hot, warm, and cold phases) since the 1970s, how magnetic fields differ between the ISM phases is still unknown. Using results from numerical simulations, this work explores how the multiphase nature of the ISM shapes magnetic fields and then discusses possible implications of those results for polarisation observations of the Milky Way and high-redshift galaxies. These findings will enhance our understanding of the role of magnetic fields in galaxy evolution and prepare us to harness the upcoming wealth of radio polarisation data from the Square Kilometre Array and its pathfinders.
\end{abstract}
\section{Introduction} \label{sec:intro}
Magnetic fields are a dynamically important component of the interstellar medium (ISM) of star-forming galaxies, most notably they reduce the efficiency of star formation \citep{FederrathK2012, KrumholzF2019, PattleEA2023} and control the propagation of cosmic rays \citep{ShukurovEA2017, SetaEA2018, RuszkowskiP2023}. The interstellar medium of galaxies is known to be a multiphase medium from the 1970s \citep{FieldEA1969, McKeeO1977} with the thermal gas at a range of densities and temperatures and the ISM is broadly divided into cold, warm, and hot phases \citep{Cox2005, Ferriere2020, McClure-GriffithsEA2023}. It is still being determined how the properties of magnetic fields differ between the ISM phases, which this work aims to discuss.

Moreover, the magnetic fields in the early Universe and protogalaxies are known to be significantly weaker \citep[$\approx 10^{-10}\,\muG$, see][]{Subramanian2016} than the present day observed strengths \citep[$\approx 5\,\muG$, see][]{Beck2016}. This amplification is due to the {\it turbulent dynamo} mechanism \citep{RuzmaikinEA1988, BrandenburgS2005, ShukurovS2021}, which is the conversion of the kinetic energy of turbulence to magnetic energy.  This process has been historically studied using numerical simulations of isothermal plasmas \citep{MeneguzziEA1981, SchekochihinEA2004, HaugenEA2004, FederrathEA2011, BhatS2013, FederrathEA2014, Cho2014, Federrath2016, SetaEA2020, SetaF2021dyn, SurS2024} for proper comparisons with analytical theories and physical models \cite[e.g.][]{Kazantsev1968, Moffatt1978, RuzmaikinEA1988, Subramanian1999, SchekochihinEA2002, Subramanian2003, Schekochihin2022}. However, applying these results directly to magnetic fields in the multiphase ISM is difficult and thus requires exploring the turbulent dynamo in a multiphase medium. This is recently done with simulations \citep{SetaF2022, GentEA2023}. The current work further discusses the turbulent dynamo process in a two-phase (cold and warm) medium and related possible observational implications.

With the motivation to study magnetic field properties and the turbulent dynamo in the multiphase ISM, this work primarily presents some of the results from \citet{SetaF2022} in \Sec{sec:numres}. Then it discusses the probable implications of those results for observational studies in \Sec{sec:obs}. Finally, we summarise our work and conclude in \Sec{sec:conc}.

\section{Numerical results} \label{sec:numres}

\subsection{Numerical setup} \label{sec:setup}
To study the properties of the turbulent dynamo and magnetic fields in a multiphase medium, we use a highly modified version of the FLASH code \citep{DubeyEA2008, WaaganEA2011, FederrathEA2021} to simulate a small patch of size $200\,\pc$ of a Milky Way-type galaxy on a three-dimensional, triply periodic domain sampled on a uniform grid with $512^{3}$ grid points. The magnetohydrodynamic equations \citep[Eq.~1 -- 4 in][]{SetaF2022} are solved with turbulence being continuously driven solenoidally \citep{FederrathEA2010, FederrathEA2022} on a scale of $\ell_{0} = 100\,\pc$ with a velocity dispersion of $u_{0} = 10\,\km \, \s^{-1}$ and employing heating and cooling functions suitable for a Milky Way-type spiral galaxy \citep[more details of the heating and cooling functions in][]{KoyamaI2000, KoyamaI2002,Vazquez-SemadeniEA2007}. Physical viscosity and resistivity (with values greater than the corresponding numerical ones) are prescribed, which leads to both the hydrodynamic and magnetic Reynolds numbers of $2000$. Initially, a uniform density of $1\,\g\,\cm^{-3}$ and temperature of $5000\,\K$ is prescribed with zero velocity and weak random magnetic field of strength $10^{-4}\,\muG$ \cite[as long as it is weak, the initial field structure do not alter the properties of the turbulent dynamo and dynamo amplified magnetic fields, see][]{SetaF2020}. The simulation is run till the magnetic field achieves a statistically steady state, which is roughly $100\,t_{\rm 0} \approx 1\,\Gyr$, where $t_{0} = \ell_{0} / u_{0} \approx 10\,\Myr$ is the eddy turnover time. Further details about the setup are given in Sec.~2 of \citet{SetaF2020}.

\subsection{Turbulent dynamo in two-phase simulations}  \label{sec:turbdyn}

The simulated ISM exhibits a two-phase nature with two dominant peaks in density and temperature distributions. We separate the medium into phases using temperature, $T$, only. Gas with $\cold$ is categorised as the cold medium and gas with $\warm$ as the warm medium. The phases show properties expected from observations \citep{HeilesT2003II, GaenslarEA2011, SchneiderEA2013, NguyenEA2019, SetaF2021pul, MarchalM2021}: the cold ISM occupies only $\approx 5\%$ of the volume, whereas the warm ISM occupies the rest $\approx 95\%$ and the turbulence is supersonic in the cold ISM (turbulent Mach no., $\Mach \approx 5$) but transonic in the warm ISM ($\Mach \approx 1$). After examining the thermodynamic and turbulent properties of the two-phase medium, we now discuss the properties of the turbulent dynamo and magnetic fields in a two-phase medium.

\begin{figure}[h]
\includegraphics[width=\columnwidth]{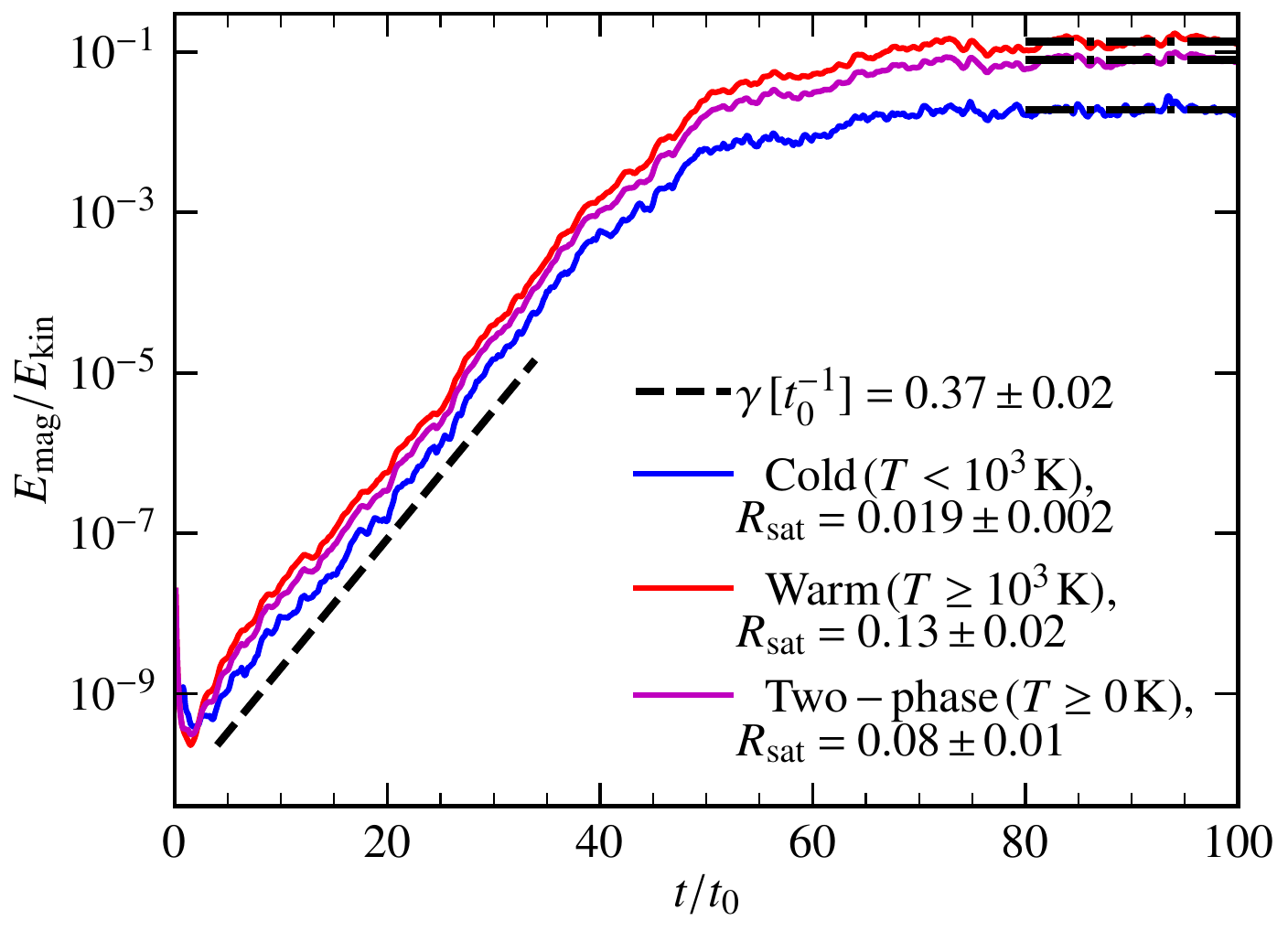} 
\caption{The evolution of the ratio of magnetic to kinetic energy, $\Emag/\Ekin$, as a function of time normalised by the eddy turnover time of the turbulence, $t/t_{0}$, for the cold ($\cold$, blue) and warm ($\warm$, red) ISM as well as the two-phase medium as a whole ($\whole$, magenta). For all three cases, the magnetic field amplifies exponentially and then saturates. The growth rate, $\gamma\approx0.37 t^{-1}_{0}$, computed by fitting a line in a log-linear space is roughly the same (dashed black line) for both the cold and warm ISM. The saturation level, $R_{\rm sat}$, obtained by fitting a horizontal line to $\Emag/\Ekin$ (dashed, dotted black line) in the saturated stage of the turbulent dynamo ($t/t_{0} \gtrsim 80$) is lower in the cold ISM ($\approx 0.019$) in comparision to the warm ISM ($\approx 0.13$).}
\label{fig:ts}
\end{figure}

\Fig{fig:ts} shows the evolution of the ratio of magnetic to turbulent kinetic energy density, $\Emag/\Ekin$, over the entire period of the simulation run, $100\,t_{0}$. Given that the turbulence is driven continuously, the kinetic energy remains statistically constant throughout the run. The weak magnetic energy at $t/t_{0} = 0$, after a small decay at the start, amplifies exponentially. This is the kinematic stage of the turbulent dynamo. Once the magnetic field becomes strong enough ($t/t_{0} \approx 40, \Emag/\Ekin \approx 10^{-3}$), the exponential growth slows down and the turbulent dynamo enters a slower, power-law growth stage. Finally, the magnetic energy achieves a statistically steady state ($t/t_{0} \gtrsim 80$) and this is referred to as the saturated stage of the turbulent dynamo. Both the cold and warm ISM phases show all three stages of the turbulent dynamo.

The estimated growth rate in the kinematic stage of the turbulent dynamo, $\gamma \approx 0.37\,t_{0}^{-1}$, is the same in both the cold ($\Mach \approx 5$) and warm ($\Mach \approx 1$) phases and this is in direct contrast with isothermal simulations which shows a lower growth rate for supersonic turbulence in comparision to transonic turbulence \citep{FederrathEA2011, SetaF2021dyn, AchikanathChirakkaraEA2021}. Thus, the turbulent dynamo in a two-phase, non-isothermal medium cannot be considered as a combination of two separate single-phase, isothermal media and this is due to significant cross-talk between the phases.  Moreover, the equal growth rate in both phases can be explained by the roughly equal generation of vorticity in each phase \citep[see Sec.~4.2 in][for details]{SetaF2022}. Thus, the turbulent dynamo is equally efficient in exponentially amplifying magnetic fields in both the cold and warm ISM phases.

Next, another quantity of interest is $\Emag/\Ekin$ in the saturated stage of the turbulent dynamo, $R_{\rm sat}$, which quantifies the total dynamo-amplified magnetic energy in terms of the turbulent kinetic energy. $R_{\rm sat}$ is significantly lower in the cold ISM phase ($\approx 0.02$) compared to the warm ISM phase ($\approx 0.13$). This result qualitatively agrees with isothermal turbulent dynamo simulations but there are still significant quantitative differences \citep[see Table~1 in][]{SetaF2022}. Physically, the lower saturation level,  $R_{\rm sat}$, in the cold ISM is due to a stronger Lorentz force leading to a stronger back reaction of the magnetic fields on the gas in comparision to the warm ISM \citep[see Sec.~4.2 in][for further details]{SetaF2022}. The difference in the Lorentz force is directly connected to the difference in the local structure of the magnetic fields, which we quantify using the magnetic field line curvature in \Sec{sec:curb}.

\begin{figure}[h]
\includegraphics[width=\columnwidth]{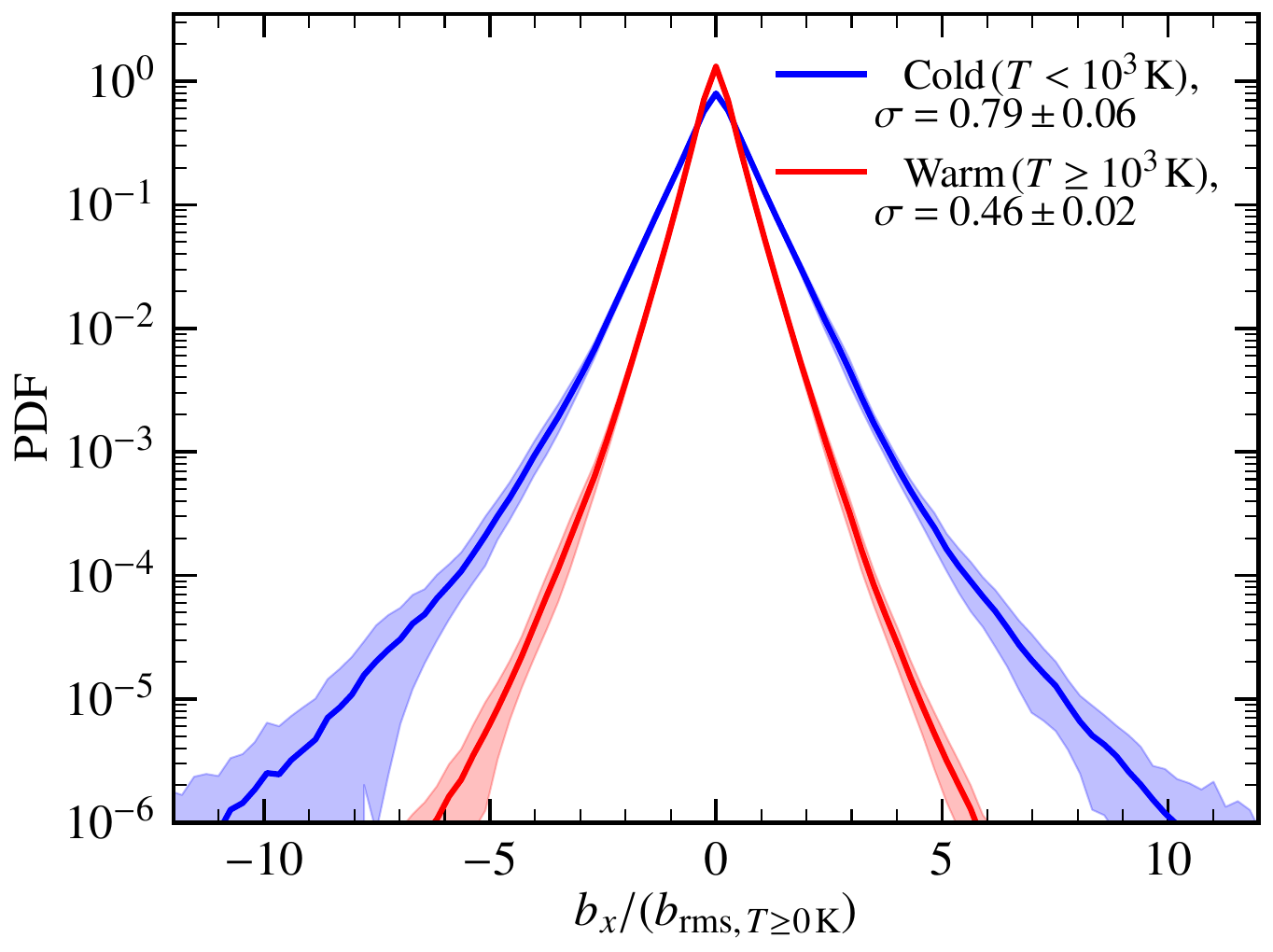} 
\caption{Probability distribution function (PDF) of the $x$-component of the magnetic field normalised by the root-mean-square (rms) strength over the entire numerical domain,  $b_{x} / (b_{{\rm rms},\,T\ge0\,{\rm K}})$, in the saturated stage of the turbulent dynamo for both the cold (blue) and warm (red) ISM phases. The lines and shaded regions show the mean and standard deviation of the PDFs averaged over $10\,t_{0}$ in the saturated stage ($90 \le t/t_{0} \le 100$ in \Fig{fig:ts}). The magnetic field strengths are statistically larger in the cold ISM in comparision to the warm ISM with the standard deviation (given in the legend) higher roughly by a factor of 2.}
\label{fig:pdf}
\end{figure}

A lower $R_{\rm sat}$ in the cold ISM implies that, in comparision to the warm ISM, a smaller fraction of the turbulent kinetic energy is converted to magnetic energy. However, this does not necessarily mean weaker magnetic fields in the cold ISM. \Fig{fig:pdf} shows the probability distribution function of a single component of the magnetic field in the saturated stage of the turbulent dynamo for both the cold and warm phases. The magnetic fields are statistically stronger in the cold phase compared to the warm phase.

\subsection{Magnetic field line curvature in two-phase simulations}  \label{sec:curb}

\begin{figure}[h]
\includegraphics[width=\columnwidth]{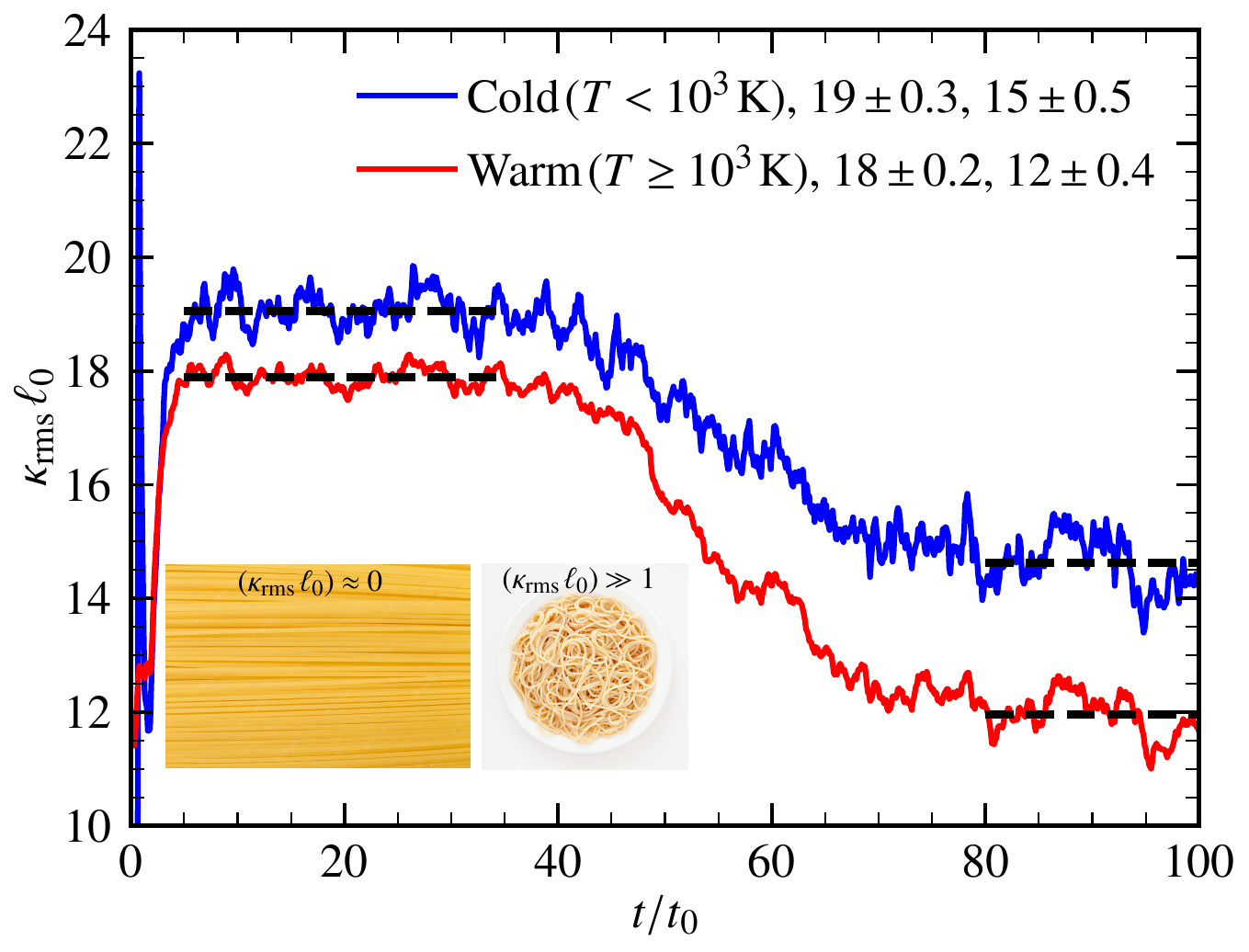} 
\caption{The evolution of root-mean-square (rms) magnetic field lines curvature normalised by the driving scale, $\kappa_{\rm rms}\,\ell_{0}$, for the cold and warm ISM. The legend shows the mean values of the quantity in the exponentially amplifying and saturated stages of the turbulent dynamo. The mathematical quantity, field line curvature, can be more simply visualised using a toy example of {\it noodles}, where flat noodles would have $\kappa_{\rm rms}\,\ell_{0} \approx 0$ and tangled ones would have $\kappa_{\rm rms}\,\ell_{0} \gg 1$. The magnetic field lines are more curved in the exponentially amplifying stage in comparision to the saturated stage. However, $\kappa_{\rm rms}\,\ell_{0}$ is always larger for the cold ISM compared to the warm ISM, implying comparatively more tangled magnetic fields in the cold phase. }
\label{fig:curb}
\end{figure}

The local structure of the magnetic field is characterised by the magnetic field line curvature, usually defined as $||\hat{b} \cdot \nabla \hat{b}||$ \citep{SchekochihinEA2004}, where $\hat{b} = \vec{b}/||\vec{b}||$ is the unit vector and $||\,||$ denotes the magnitude of the quantity. However, in numerical simulations, it is more accurately captured by a modified form, $\kappa = ||\hat{b} \times (\hat{b} \cdot \nabla \hat{b})||$ \citep{YangEA2019, YuenL2020}.

\Fig{fig:curb} shows the computed root-mean-square (rms) of the magnetic field line curvature over each phase, $\kappa_{\rm rms}$, normalised by the driving scale of turbulence, $\ell_{0}$, for both the cold and warm ISM. The magnetic field lines, for both the ISM phases, are more tangled during the kinematic stage of the turbulent dynamo in comparision to the saturated stage but the magnetic fields are always statistically more curved in the cold ISM in comparision to the warm phase.

\section{Possible observational implications}  \label{sec:obs}
We confirm that the magnetic field properties differ with the ISM phase from the numerical results in \Sec{sec:numres}. In particular, the ratio of magnetic to turbulent kinetic energy in the saturated stage of the turbulent dynamo is lower in the cold phase in comparision to the warm phase (\Fig{fig:ts}) but the magnetic field strengths are statistically higher in the cold phase (\Fig{fig:pdf}). Also, the magnetic field is statistically more tangled in the cold phase  (\Fig{fig:curb}). In this section, we discuss the possible observational implications of these results. At the onset itself, we caution that these are currently only ideas and that more work is required for a systematic comparision between numerical results and observations, which we aim to do in the future.

\subsection{Derived magnetic field strengths in high-redshift galaxies}  \label{sec:highredshift}
It is important to observationally probe magnetic fields in high-redshift galaxies to constrain turbulent dynamo theories \citep{BernetEA2008, MaoEA2017, BasuEA2018, SetaEA2021, ShahS2021}. Recently, thanks to gravitational lensing, polarised thermal emission from magnetically aligned dust grains was observed in high-redshift galaxies, 9io9 at $z=2.6$ \citep{GeachEA2023} and SPT0346-52 at $z = 5.6$ \citep{ChenEA2024}. The authors assumed energy equipartition between the magnetic and turbulent kinetic energies ($\Emag/\Ekin=1$) and derived the upper limits of the magnetic field strength to be $\approx 514\,\muG$ for 9io9 and $\approx 450\,\muG$ for SPT0346-52 \citep[strong magnetic fields of $\gtrsim 200\,\muG$ are also reported for nearby galaxies with far-infrared polarisation observations, see][]{Lopez-RodriguezEA2021, TramEA2023}. These are roughly two orders of magnitude higher than the strengths observed in nearby galaxies \citep[$\approx 5\,\muG$,][]{Beck2016}. If we assume that the emission is primarily from the cold ISM (not unreasonable since the dust grains are mostly embedded in the molecular ISM) and from our numerical results if we instead assume $\Emag/\Ekin \approx 0.02$ (\Sec{sec:turbdyn} and \Fig{fig:ts}), we obtain field strengths $\approx 70\,\muG$. This is still significantly higher than the average field strength in a typical nearby galaxy and that in another high-redshift galaxy at $z=0.439$ probed via synchrotron polarisation observations \citep[$\approx 10\,\muG$,][]{MaoEA2017}. These differences might also stem from the inherent bias in observational probes, e.g.~the far-infrared polarisation probes more of the cold ISM and synchrotron polarisation probes more of the warm ISM \citep{Martin-AlvarezEA2024} and the cold ISM might host stronger fields (\Fig{fig:pdf}) but the precise quantitative difference in field strengths between the ISM phases is yet to be determined. Moreover, field strengths of $\gtrsim 50\,\muG$ are observed in the Galactic Centre \citep{CrockerEA2010} and also at such high-redshifts, the star-formation activity \citep{MadauD2014} might be significantly higher (higher star formation rate $\rightarrow$ stronger feedback $\rightarrow$ larger levels of turbulent kinetic energy $\rightarrow$ stronger magnetic fields). Thus, these high-redshift galaxies may harbour magnetic fields with strengths significantly stronger than nearby galaxies. Still, the dependence of magnetic fields on the multiphase nature of the ISM should be considered while interpreting such multiwavelength observations. Also, we highlight that these equipartition arguments cannot explain the large magnetic field scale ($\approx 5\,\kpc$) observed in the young galaxy, 9io9 at $z=2.6$ \citep{GeachEA2023} and explaining such a large length scale at such a high-redshift is a significant challenge for current dynamo theories \citep{Beck2023}.

\subsection{Magnetic field structure in the Galaxy from correlation of multiple observations} \label{sec:struc}
For high-latitude (Galactic latitude $> 30^{\circ}$) sky and low total neutral hydrogen column density, $N_{\rm HI} < 4 \times 10^{20}\,\cm^{-2}$, the dust polarisation fraction (referred to as the variable $y$) at 353 GHz from Planck is shown to be better correlated with the fraction of cold neutral medium ($f_{\rm CNM}$) along the line of sight (first $x$ variable) than $N_{\rm HI}$  \citep[second $x$ variable, see][for further details]{LeiC2024}. From this inference and simple data-driven modelling, the authors conclude that either the magnetic field is less tangled in the cold neutral medium (CNM) in comparision to the warm neutral medium (WNM) or that the magnetic field in the CNM is comparatively less sampled because of smaller CNM scales. Their first conclusion seems to be in direct contradiction with the numerical results presented in \Sec{sec:curb}. However, significant caution is required for such a comparision. For example, in contrast to quantifying tangling using magnetic field data only in \Sec{sec:curb}, the polarisation fraction measurements are density-weighted, which introduces some non-trivial connection between the $x$ and $y$ variables of the correlation. Moreover, well-separated CNM clouds might not have their magnetic field oriented along similar directions and can have completely random orientations, which further complicates modelling.  A proper study with such observational and simulated results requires an `apples-to-apples' comparision using synthetic observations from multiphase ISM simulations, which we aim to discuss in the future (Seta et al., in prep.). Furthermore, enhancements in numerical simulations, especially including more physics (e.g.~shear due to galactic differential rotation and turbulence injected by supernova explosions), are required for more realistic modelling of the multiphase ISM.

\section{Conclusions} \label{sec:conc}
Using numerical experiments, we explored how the properties of the turbulent dynamo and magnetic fields depend on the phase of the ISM. We showed that the growth rate in the exponentially amplifying stage of the turbulent dynamo is the same in both the cold and warm phases but the saturation level (i.e.,~the ratio of magnetic to turbulent kinetic energy in the saturated stage) is lower in the cold phase in comparision to the warm phase (\Fig{fig:ts}). However, the magnetic fields are statistically stronger in the cold ISM in comparision to the warm ISM (\Fig{fig:pdf}). Furthermore, we also show that the magnetic fields are more tangled in the cold phase (\Fig{fig:curb}). These results are also discussed in the context of recent observational studies (\Sec{sec:obs}). We emphasize that results from such numerical experiments can help model the multiphase gas-magnetic field connection in galaxy- and cosmological-scale simulations to clarify the role of magnetic fields in the evolution of galaxies and better interpret fine-scale magnetic field observations from the Square Kilometre Array and its pathfinders \citep{HealdEA2020}.

\section*{Acknowledgements}
AS thanks Enrique Lopez Rodriguez, Minjie Lei, Susan Clark, and Antoine Marchal for their constructive comments and suggestions on the proceedings. AS further thanks Christoph Federrath, Naomi M. McClure-Griffiths, Yik Ki Ma, Hiep Nguyen, Cameron Van Eck, and Craig Anderson for useful discussions.  This work is funded by the Australian Research Council Discovery Projects (DP190101571, awarded to Naomi M. McClure-Griffiths and DP230102280, awarded to Christoph Federrath). AS acknowledges high-performance computing resources provided by the Australian National Computational Infrastructure (grant ek9) and the Pawsey Supercomputing Centre (project pawsey0810) in the framework of the National Computational Merit Allocation Scheme and the ANU Merit Allocation Scheme.

\bibliography{preSKA}
\bibliographystyle{mnras}

\label{LastPage}
\end{document}